\documentclass{article}
\usepackage{spconf,amsmath,graphicx}
\usepackage{caption}
\usepackage{makecell}
\mathchardef\mhyphen="2D

\usepackage{graphicx}
\usepackage{hyperref}
\usepackage{url}
\usepackage{epstopdf}

\usepackage{amsmath,amsfonts,bm}

\def\eqref#1{equation~\ref{#1}}

\def\1{\bm{1}}

\def\vu{{\bm{u}}}

\def\vx{{\bm{x}}}

\def\vz{{\bm{z}}}

\def\mS{{\bm{S}}}

\DeclareMathAlphabet{\mathsfit}{\encodingdefault}{\sfdefault}{m}{sl}
\SetMathAlphabet{\mathsfit}{bold}{\encodingdefault}{\sfdefault}{bx}{n}

\title{Unified Signal Compression Using Generative Adversarial Networks}
\name{Bowen Liu\textsuperscript{*}, Ang Cao\textsuperscript{*}, Hun-Seok Kim\thanks{\textsuperscript{*}Equally contributed first authors.}}
\address{EECS, University of Michigan, Ann Arbor, Michigan, USA}

\begin{document}
%
\maketitle
\begin{abstract}
We propose a unified compression framework that uses generative adversarial networks (GAN) to compress image and speech signals. The compressed signal is represented by a latent vector fed into a generator network which is trained to produce high quality signals that minimize a target objective function. To efficiently quantize the compressed signal, non-uniformly quantized optimal latent vectors are identified by iterative back-propagation with ADMM optimization performed for each iteration. Our experiments show that the proposed algorithm outperforms prior signal compression methods for both image and speech compression quantified in various metrics including bit rate, PSNR, and neural network based signal classification accuracy.
\end{abstract}


\section{Introduction}
\label{sec:intro}



 In the era of big data, signal compression is a task of critical importance to minimize the network bandwidth on the channel shared between many edge devices. Signal compression also significantly reduces the energy overhead of wireless data communication, which often dominates the overall energy consumption of power-constrained Internet of Things devices. Inspired by recent remarkable success of generative adversarial networks (GAN) in various applications, we propose a unified signal compression framework called BPGAN (back propagated GAN) where the compressed signal is represented by a latent vector fed into a generator network which is trained to produce realistic high quality signals. The core idea of BPGAN is to `search' an optimal latent vector through iterative back-propagation for a given generator (with fixed weights) and the target signal to be compressed. This process minimizes a loss function computed based on the generator output and the target signal, enabling high quality compressed signal represented by the latent vector input to the generator. This framework is generally applicable to different type of signals including speech and image as long as a GAN is trainable in that signal domain.
\section{Related Work}
\label{relatedwork}

Deep auto-encoder (DAE) and deep neural network (DNN) based image compression has been demonstrated in literature including \cite{balle2016end} and \cite{rippel2017real}. The main idea of these prior work is to train a DNN based encoder and decoder pair that optimizes mean squared error (MSE) and/or other metric such as multi-scale structural similarity (MS-SSIM) for image quality assessment between the original and decoder output images.

GAN was  introduced by \cite{goodfellow2014generative}, where a generator is trained to generate realistic but synthesized images and a discriminator is trained to distinguish real vs. synthesized images. The generator and discriminator in a GAN are trained sequentially to outperform each other. Recently, \cite{mentzer2018conditional} and \cite{agustsson2018generative} applied GAN based training to deep (auto-)encoder networks to compress images with extremely low data rates with aesthetically pleasing details on the generated image. However, these `realistic' details generated by the decoder (or generator) often fail to capture the `actual' details of the original image.

Traditional speech codec methods such as CELP \cite{schroeder1985code}, Opus\cite{valin2012definition}, and adaptive multirate wideband (AMR-WB) \cite{bessette2002adaptive} commonly employ hand-engineered encoder-decoder pipelines relying on manually/mathematically crafted audio representation features and/or signal prediction models. Recent DNN based approaches including \cite{kankanahalli2018end} demonstrate the feasibility to train an end-to-end speech codec that exhibits performance comparable to a hand-crafted AMR-WB codec at 9-24 kbps. \cite{cernak2016composition} uses deep spiking neural networks to realizes a low bit rate speech codec. Another strategy to realize a high quality speech codec is to use a DNN based vocoder such as Wavenet \cite{oord2016wavenet} and WaveRNN\cite{kalchbrenner2018efficient} as a decoder to synthesize speech. These methods, however, do not scale well to a very low bit-rate (e.g, 2kbps). We propose BPGAN for speech compression to overcome limitations in prior work.

\section{BPGAN Compression Framework}
\label{framework}
The BPGAN  compression is applicable to any signal type as long as it is possible to train a GAN that produces realistic generated outputs in that type. The overall flow of the  BPGAN compression is shown in Figure 1. Unlike other GAN based approaches that rely on an encoder \cite{agustsson2018generative} that provides a compressed signal, compression in BPGAN is performed by iteratively searching and updating the generator input latent vector through back-propagation to minimize a target loss function at the output of the generator. The optimal latent vector input for the generator is the compressed signal as shown in Figure 1. Selecting a proper loss function and iteratively updating/searching the input of the generator through back-propagation is the key step to significantly improve the quality and/or compression ratio of the signal. As this compression framework allows applying various loss functions during iterative back-propagation, it enables objective-aware signal compression to obtain the optimized compression results tailored for a target application such as signal classification and recognition. The iterative back-propagation search process is combined with alternating direction method of multipliers (ADMM)-based non-uniform quantization and Huffman coding to further reduce the size of the latent vector (i.e., compressed signal). To accelerate this iterative back-propagation based compression process, we initialize the latent vector using the output of an encoder as an initial latent vector. It reduces the number of iterations from 2500 to 200 -- 550. In the decompression stage, the compressed signal (i.e., the latent vector input for the generator) is fed into the same generator to (re-)generate the decompressed signal.

\begin{figure}[h]
\label{Frameworkoverview}
\begin{center}
\vspace{-5mm}
\includegraphics[width=0.5\textwidth]{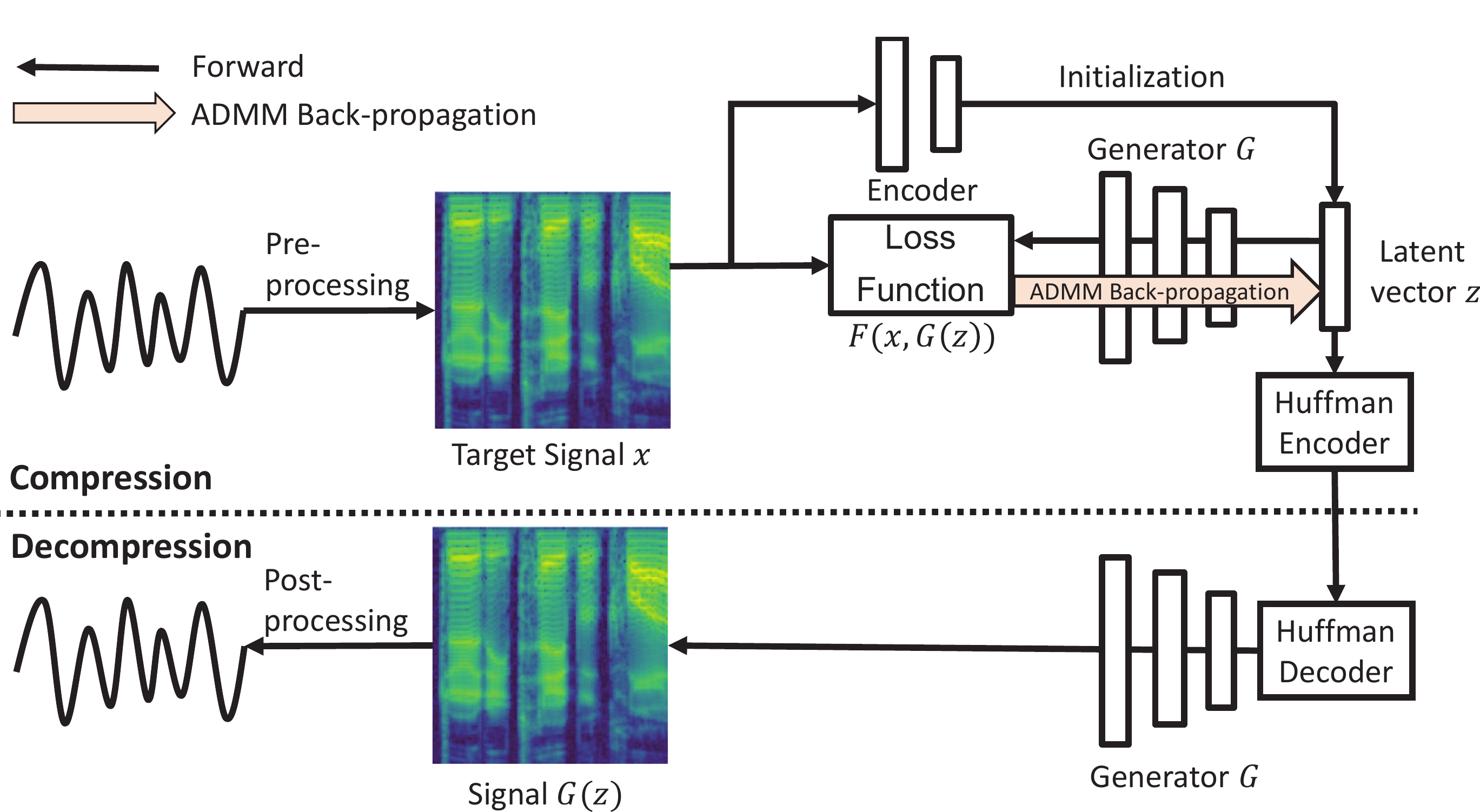}
\end{center}
\vspace{-5mm}
\caption{BPGAN Compression Framework Overview}
\vspace{-5mm}
\end{figure}

\subsection{BPGAN Training Methodology}

Our BPGAN training methodology consists of two stages:

 \textbf{Stage one}: Training a GAN with floating point input for the target signal type. This step is exactly the same as a typical GAN training procedure in \cite{wang2018high}, where a generator $G$ and discriminator $D$ are adversarially trained. An encoder can be cascaded by the generator to form an auto-encoder structure, where the encoder is trained to learn mappings from the signal to a latent space vector. 

 \textbf{Stage two}: Training a GAN with quantized input. After completing Stage one, we perform signal compression on the training set images through iterative back-propagation and quantization. Then we retrain the GAN by only using quantized latent vectors to improve performance of the GAN in the quantized latent vector space.

 When the training is finished, the weights of the GAN are frozen for signal compression and decompression. The iterative back-propagation process for the optimal compressed vector search only updates the latent vector, not the weights of the GAN.

\subsection{Compression methodology details}

Given a well-trained GAN with a generator $G$, the compression procedure \textit{searches} for the optimal input latent vector $\vz$ of $G$ through iterative back-propagation to generate an output $G(\vz)$ that minimizes an appropriate loss function $F(\vx,G(\vz))$ evaluated with the original target signal $\vx$. This compression process can be expressed as an optimization problem:
\begin{equation}
    \hat{\vz} = \arg\min_{\vz} F(\vx,G(\vz))
    \label{eq:opt}
\end{equation}

The dimension of $\vz$ is much smaller compared with that of the original $\vx$, thus the signal is compressed. We demonstrate in the evaluation section that by making use of appropriate loss function and efficient quantization method, this framework is capable of achieving higher compression ratio and/or better quality signal at the same compression ratio compared to the state-of-the-art techniques. The optimization problem (\ref{eq:opt}) is solved by the iterative back-propagation process based on the gradient $\nabla F_{\vz}$ where $\vz$ is initialized by the encoder output and quantized at every iteration.


Quantization of $\vz$ is essential to further reduce the number of bits for a higher compression ratio. Thus, we formulate the latent vector search problem as an optimization problem with a quantization constraint and obtain a solution based on ADMM \cite{wei2012distributed}. That is, given $G$, $\vx$, and $F(\vx,G(\vz))$, the problem of finding an optimal quantized latent vector $\vz$ is formulated as:
\begin{equation}
    \arg\min_{\vz,\vu} F(\vx,G(\vz)) + I(\vu \in \mS) \quad s.t. \vu=\vz
\end{equation}
where $\mS$ is a non-convex set whose elements are all quantized vectors, $\vu$ is an auxiliary variable and $I(\cdot)$ is an indicator function. This optimization problem with non-convex constraints is difficult to solve directly, therefore we rewrite this equation and apply ADMM to solve it.
The augmented Lagrangian of the above optimization problem is given by:
\begin{multline}
    L(\vz,\vu,\bm{\eta},\mu) = F(\vx,G(\vz)) + \\ I(\vu \in \mS) + \frac{\mu}{2}(\|\vz - \vu + \bm{\eta}) \|^{2}_{2} - \|\bm{\eta}\|^2_2)
\end{multline}
ADMM is designed to minimize $ L(\vz,\vu,\bm{\eta},\mu)$ by updating variables $\vz, \vu, \bm{\eta}$ alternatively in every iteration. The ADMM updating procedure for $k=0,1,2,\dots$ is given by:
\begin{gather}
    \vz_{k+1} =  \arg \min_{\vz} F(\vx,G(\vz)) + \frac{\mu}{2}\|\vz - \vu_{k} + \bm{\eta}_{k}\|^2_2
   \\
   \vu_{k+1} = \arg \min_{\vu} I(\vu \in \mS) + \frac{\mu}{2}\|\vz_{k+1} - \vu + \bm{\eta}_{k}\|^2_2 \\
   \bm{\eta}_{k+1} = \bm{\eta}_{k} + \vz_{k+1} - \vu_{k+1}
\end{gather}
$\vz$ updating steps are typical back-propagation with additional term of $L_2$ regularization. For $\vu$ updating steps, the solution is $\vu_{k+1} = Q(\vz_{k+1}+\bm{\eta}_{k})$ where $Q(\cdot)$ is a non-uniform quantization function which directly project into the set of quantized vectors $S$. The non-uniform quantization centers of $S$ are obtained by K-means clustering based on the distribution of unquantized latent vectors. After ADMM process, we apply quantization $Q$ directly on the latent vectors to ensure quantized input, which repeats until the loss function is minimized. Finally we apply lossless Huffman coding on the quantized latent vectors to further compress the signal.


\section{Experiment}
We test our BPGAN compression on image and speech compression tasks using the generator structure in Fig. 2. 
\begin{figure}[h]

\begin{center}
\vspace{-2mm}
\includegraphics[width=0.485\textwidth]{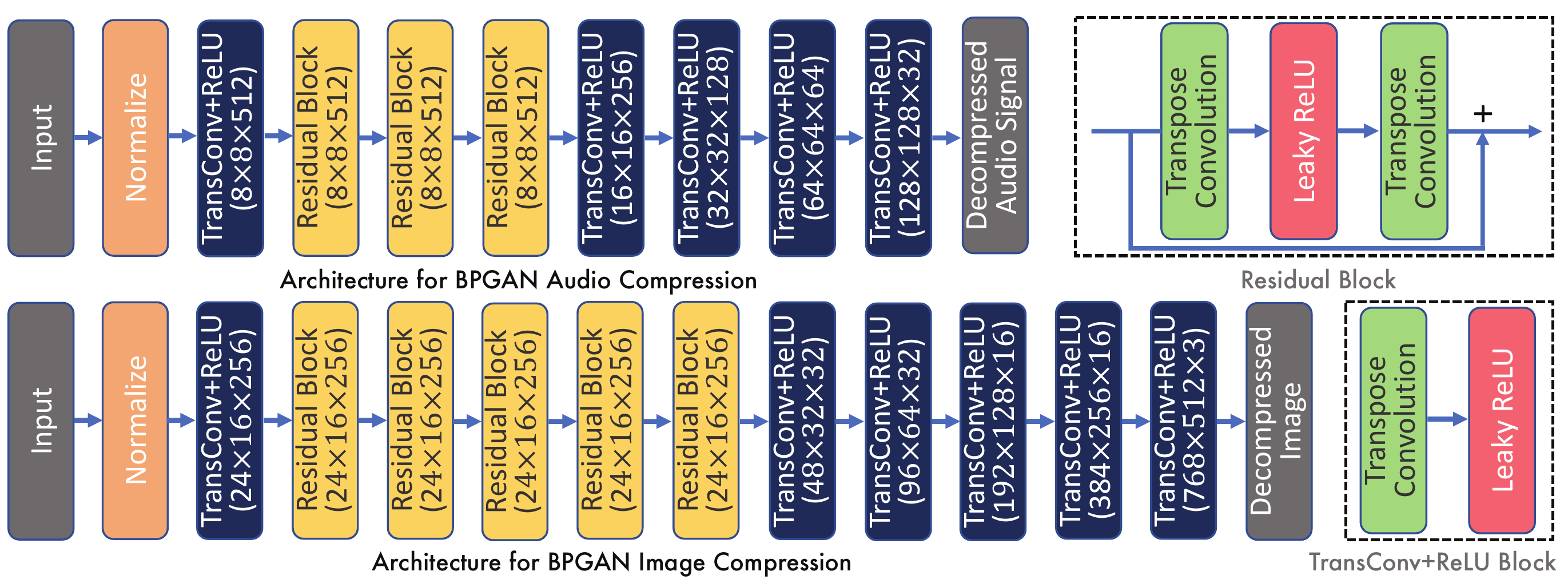}
\end{center}
\vspace{-5mm}
\caption{BPGAN generators for image and speech compression with transpose convolution layers and residual blocks.}
\label{Arch}
\vspace{-5mm}
\end{figure}
\vspace{-3mm}
\subsection{Image Compression Evaluation Setup}
BPGAN image compression is trained and evaluated with three benchmark datasets specified in the next subsection. We rescale images from all datasets for the uniform size input of 768$\times$512 pixels.  Compression performance is quantified with objective metrics such as PSNR and MS-SSIM, which are widely used but known to be often unreliable to capture the subjective image quality. Hence, we include ImageNet classification results to quantify the image quality.

The Open Images Dataset V5 \cite {OpenImages} containing 9M images is used for the GAN network training. The Kodak dataset \cite {Kodak} is used for image compression to evaluate the PSNR and MS-SSIM results. The ImageNet is used to measure the classification accuracy when the decompressed images are classified by a VGG-16 network \cite {simonyan2014very}. Note that the VGG-16 network is trained with uncompressed original images only.

For the back-propagation loss function $F$ for image compression, we combine MSE and MS-SSIM loss  \cite{zhao2016loss}:
\begin{equation*}
    F(\vx,G(\vz) =   \mathcal{L}^{\mathbf{MS\mhyphen SSIM}} (\vx,G(\vz))
  +\alpha \cdot MSE(\vx, G(\vz))
\end{equation*}

The baselines for our method comparison are BPG\cite{BPG}, JPEG, and another GAN based method  \cite{agustsson2018generative}. 

\subsection{Audio Compression Evaluation Setup}

For speech compression, we use the log-magnitude of mel-spectrograms \cite{marafioti2019adversarial} of speech as the input for compression instead of using the time domain speech directly.  
Our experiments are carried on the TIMIT dataset \cite{garofolo1993darpa}, which contains a total of 6300 sentences spoken by 630 speakers from 8 major dialect regions of the United States at a sample rate of 16 kHz. We divide the dataset into the training and testing subset that are strictly disjoint.

For speech spectrogram representations, we use short-time Fourier transform (STFT) with 128-sample stride and 512-sample frame size to compute magnitudes, resulting in 75\% inter-frame overlap. 
Then we transform spectrograms into mel-spectrograms with 128 mel-frequency bins for each frame and collect 128 frames to construct a 128$\times$128 mel-spectrogram that corresponds to one second speech audio. The normalized log-magnitude of mel-spectrograms in the range of $[-1,1]$ is the BPGAN compression input $\vx$.  To reconstruct the time domain speech from decompressed mel-spectrograms, we employ  phase-gradient heap integration (PGHI)
\cite{pruuvsa2017noniterative} method followed by inverse STFT. 


The loss function used in speech compression is defined as
\begin{equation*}
    F(\vx,G(\vz) = \mathcal{L}^{\mathbf{feat}} (\vx,G(\vz)) + \alpha  \cdot MSE(\vx, G(\vz))
\end{equation*}
where $\mathcal{L}^{\mathbf{feat}}$ measures the L2 distance between feature maps from convolution layers of a VGG-BLSTM \cite{liu2019adversarial} network using generated and original speech signals as input. We evaluate the quality of the compressed signal using a phoneme recognition task performed on TIMIT dataset using the joint CTC-attention loss \cite{kim2017joint}. 


\begin{figure*}
\centering
\includegraphics[width=175mm]{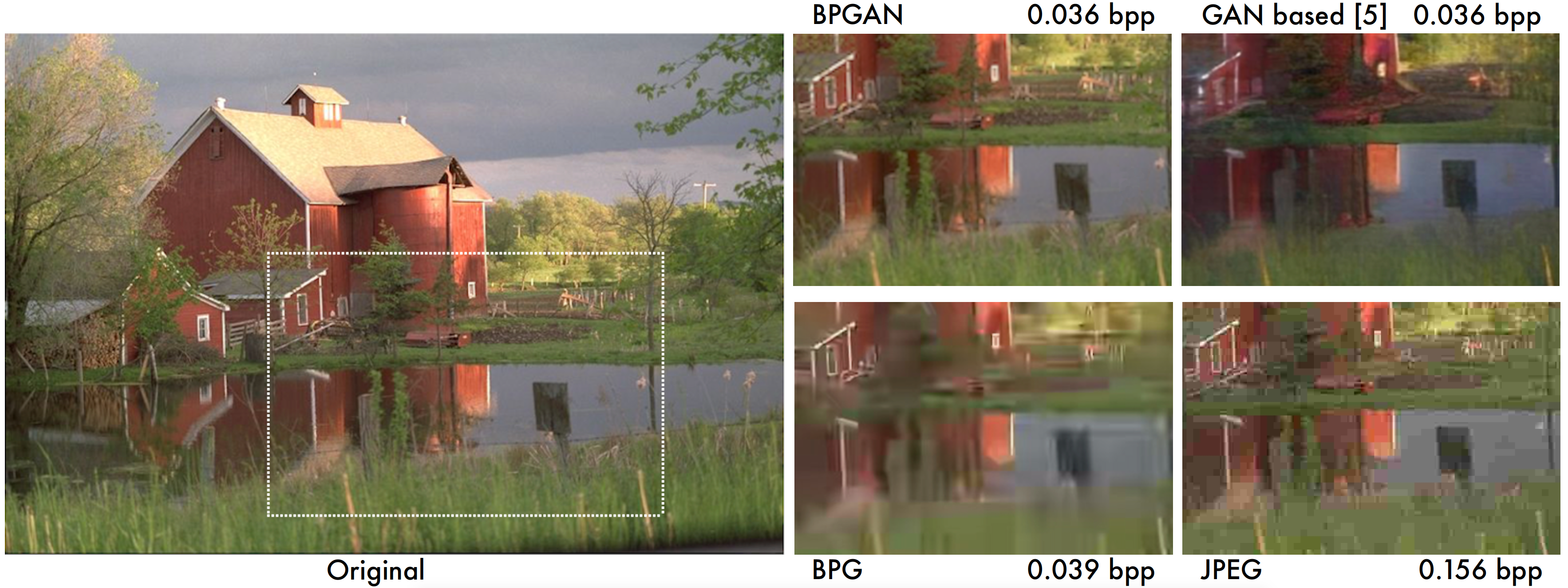}
\vspace{-5mm}
\caption{Visual comparison on Kodak: BPGAN outperforms other methods with similar bpp preserving the original details.}
\vspace{-5mm}
\label{fig:picture001}
\end{figure*}

\subsection{Evaluation Results}

We summarize our experimental results in Table \ref{result_table}. PSNR and MS-SSIM evaluation is performed on Open Image V5 and Kodak datasets with the scaled image size of 768$\times$512 pixels with RGB representation (24-bit per pixel, bpp). The compressed image size of 0.286bpp is obtained by using a latent vector $\vz$ size of 20000 and 64 non-uniform quantization levels for each element after Huffman coding. For classification evaluation, we use ImageNet dataset with the original resolution of 256$\times$256 pixels. For speech compression, the original speech is sampled at 16k samples per second with 16-bit per sample. The compressed speech rate of 2 kilo-bits per second (kbps) is achieved by using a latent vector $\vz$ size of 512 and 16 non-uniform quantization levels for each element.

\begin{table}[ht]
\caption{Signal Compression Results}
\vspace{-2mm}
\centering
\resizebox{\columnwidth}{!}{
\begin{tabular}{|c|c|c|c|c|c|c|}
\Xhline{1pt}

\thead{Image \\ Methods} & \thead{Bitrate\\(bpp)} & \thead{PSNR} & \thead{MS-SSIM} & \thead{ImageNet \\Top-1 error\%} & \thead{ImageNet\\Top-5 error\%} &
\\
\Xhline{1pt}
Original & 24 & - & - & 23.7 & 6.8 &\\
\bf BPGAN & \bf 0.286 & \bf 32.9 & \bf 0.968 & \bf 23.7 & \bf 6.8 &\\
GAN based \cite{agustsson2018generative} & 0.305 & 28.2 & 0.922 & 26.0 & 7.9 &\\
JPEG & 0.306 & 26.9 & 0.864 & 42.5 & 16.6 &\\
BPG & 0.298 & 32.3 & 0.961 & 25.8 & 7.4 &\\
\Xhline{1pt}

\thead{Speech \\ Methods} & \thead{Bitrate\\(bps)} & \thead{PESQ} & \thead{MUSHRA} & \thead{Kaldi\\PER\%} & \thead{MLP\\PER\%} & \thead{LSTM\\PER\%}
\\
\Xhline{1pt}
Original & 256k       & 4.50 & 95.0  &18.7 &18.6 &15.4 \\
\bf BPGAN & \bf 2k       & \bf 3.25 & \bf 64.1  &\bf 20.9 &\bf 20.8 &\bf 18.6 \\
    CELP &4k        & 2.54 & 32.0  &28.2 &27.6 &27.3 \\
    CELP &8k        & 3.39 & 59.4  &23.0 &23.6 &21.2 \\
    Opus &9k        & 3.47 & 79.3  &22.7 &23.7 &21.2 \\
    AMR  &6.6k      & 3.36 & 58.9  &22.6 &23.6 &22.3 \\
\Xhline{1pt}
\end{tabular}
}
\label{result_table}
\vspace{-5mm}
\end{table}

Table~\ref{result_table} confirms that the proposed BPGAN compression method produces higher quality images measured in PSNR, SSIM, and ImageNet classification accuracy for the similar compressed rate of $\approx 0.3$bpp compared to other prior compression methods. Notice significant ImageNet classification accuracy difference between the proposed BPGAN (lossless) and others. Subjective compressed image quality comparison can be seen in Figure 2. It is worth noting that a GAN based method \cite{agustsson2018generative} synthesizes `realistic' detail in the compressed image but it is often inaccurate representation of the actual detail which is better preserved in the BPGAN compression output.




In order to assess our speech compression quality, we use both subjective and objective metrics. PESQ \cite{rix2001perceptual} is a metric ranging from -0.5 to 4.5 produced by an algorithm that predicts the subjective mean opinion score (MOS) of speech. We also conducted a subjective evaluation with 10 users to provide a score in Multiple Stimuli with Hidden Reference and Anchor (MUSHRA) \cite{vincent2006preliminary} standard (higher the better). In addition, we perform phoneme recognition tests to measure the phoneme error rate (PER, lower the better) using the combination of SGMM and Dans DNN in Kaldi \cite{povey2011kaldi}, and also using MLP and LSTM networks \cite{ravanelli2019pytorch} which take MFCC as the input. These phoneme recognition models are trained with original audios without any compression.





Speech compression evaluation results are summarized in the bottom half of Table~\ref{result_table}. While the proposed BPGAN based compression provides the lowest data rate of 2 kbps, it exhibits a better MUSHRA subjective quality score than other methods with higher data rates except for Opus at 9 kbps (4.5$\times$ higher data rate than ours). We have observed that while the PESQ scores are similar among multiple methods, it does not accurately predict the quality of speech metric. The PER measured by phoneme recognition tests indicates that the error rate for the proposed BPGAN is significantly lower while providing the lowest data rate compared to other methods.



Finally, in Figure 4, we show the tradeoff space between the loss of compressed speech quality (x-axis, lower the better) vs. the achievable bit rate (y-axis, without Huffman coding, lower the better). The gain of ADMM based non-uniform quantization is also shown in the same figure. One can notice that along the pareto-optimal line, combination of the optimal latent vector dimension and the number of quantization levels per element changes for various rate - quality target points. 


\begin{figure}[h]
\vspace{-4mm}
\begin{center}
\label{Audioparameterfigure}
\includegraphics[width=0.4\textwidth]{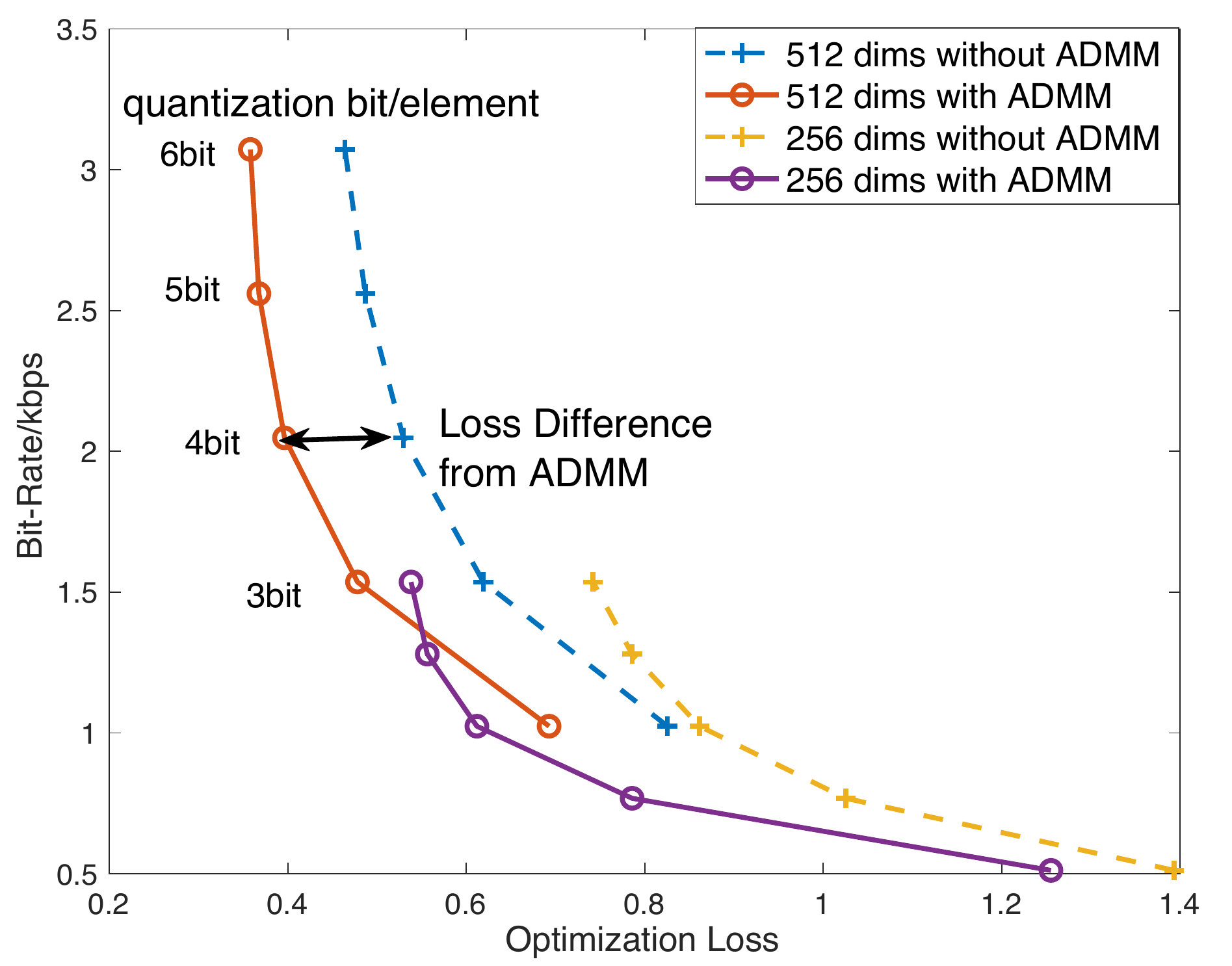}
\vspace{-5mm}
\end{center}
\caption{Parameter sensitivity evaluation: Tradeoff between the rate and quality is obtained by  adjusting the vector size and number of quantization levels for each element.  }
\vspace{-5mm}
\end{figure}

\section{Conclusion}
In this paper, we demonstrated a unified GAN based signal compression framework for image and speech signals. The proposed BPGAN compression method utilizes iterative back-propagation to search the optimal latent vector that minimizes a loss function that quantifies the compressed signal quality. Experiment results confirm that BPGAN compressed signal exhibits significantly lower data rate and/or better signal quality compared to other methods evaluated with various metrics including neural network based signal classification.

\vfill\pagebreak
\bibliographystyle{IEEEbib}
\bibliography{refs}

\end{document}